\documentclass[manuscript]{acmart}

\AtBeginDocument{%
  }

\usepackage{url}
\usepackage{soul}
\usepackage{xcolor}
\usepackage[T1]{fontenc}

\newboolean{showChanges}
\setboolean{showChanges}{False} 

\ifthenelse{\boolean{showChanges}}{%
\newcommand{\newtext}[1]{{\textcolor{magenta}{#1}}} 
\newcommand{\oldtext}[1]{\sout{#1}}
}{
\newcommand{\oldtext}[1]{}
\newcommand{\newtext}[1]{#1} 
}

\usepackage{subcaption}
\usepackage{setspace} 
\usepackage{booktabs}
\usepackage{siunitx}
\usepackage{makecell}
\usepackage{tabularx}
\begin{document}
\title{The Role of Inclusion, Control, and Ownership in Workplace AI-Mediated Communication}


\author{Kowe Kadoma}
\email{kk696@cornell.edu}
\affiliation{%
  \institution{Cornell University}
  \city{Ithaca}
  \state{New York}
  \country{USA}
}

\author{Marianne Aubin Le Qu\'{e}r\'{e}}
\email{msa258@cornell.edu}
\affiliation{%
  \institution{Cornell University}
  \city{Ithaca}
  \state{New York}
  \country{USA}
  }

\author{Xiyu Jenny Fu}
\email{xf89@cornell.edu}
\affiliation{%
  \institution{Cornell University}
  \city{Ithaca}
  \state{New York}
  \country{USA}
}

\author{Christin Munsch}
\email{christin.munsch@uconn.edu}
\affiliation{%
 \institution{University of Connecticut}
 \city{Storrs}
 \state{Connecticut}
 \country{USA}
 }

\author{Dana\"{e} Metaxa}
\email{metaxa@seas.upenn.edu}
\affiliation{%
  \institution{University of Pennsylvania}
  \city{Philadelphia}
  \state{Pennsylvania}
  \country{USA}
  }

\author{Mor Naaman}
\email{mor.naaman@cornell.edu}
\affiliation{%
  \institution{Cornell Tech}
  \city{New York}
  \state{New York}
  \country{USA}
  }

\renewcommand{\shortauthors}{Kadoma et al.}

\begin{abstract}
\oldtext{Large language models (LLMs) can exhibit social biases. 
Given LLMs’ increasing integration into workplace software, these biases may impact workers’ well-being and may disproportionately impact minoritized communities. 
This short paper investigates how co-writing with an LLM impacts three measures related to user’s well-being: feelings of inclusion, control, and ownership over their work.}
\newtext{Given large language models’ (LLMs) increasing integration into workplace software, it is important to examine how biases in the models may impact workers.  
For example, stylistic biases in the language suggested by LLMs may cause feelings of alienation and result in increased labor for individuals or groups whose style does not match.
We examine how such writer-style bias impacts inclusion, control, and ownership over the work when co-writing with LLMs.}
In an online experiment, participants wrote hypothetical job promotion requests  using either hesitant or self-assured auto-complete suggestions from an LLM and reported their subsequent perceptions. 
We found that the style of the AI model did not impact perceived inclusion. 
However, individuals with higher perceived inclusion did perceive greater agency and ownership, an effect more strongly impacting participants of minoritized genders. 
Feelings of inclusion mitigated a loss of control and agency when accepting more AI suggestions. 
\oldtext{Future work should explore feelings of inclusion in AI-written communication.}
\end{abstract}




\maketitle

 \section{Introduction}

\oldtext{Large language models (LLMs) have the potential to transform workplace communication. 
These models are already seeing widespread use assisting in common workplace tasks like writing emails. 
Recent scholarship has shown that workers can use LLMs at work to increase their productivity and produce higher-quality work~\cite{Zhang-2023-productivity}. 
However, as others have noted, one potential challenge is that language models can reflect social biases in the language they produce~\cite{weidinger-etal-2022-risks,Bender-etal-2021-parrots}. 
While the existence of such biases is well-documented, we know far less about how these biases affect individual users. 
In particular, when AI tools—which may reflect prevailing social norms—are used in the workplace, they may inflict a ``psychological tax’’ for individuals who do not align with these norms~\cite{weidinger-etal-2022-risks}, affecting their perceived inclusion, control, and ownership over their work output.}

\oldtext{Feelings of inclusion, control, and ownership are core to the human experience~\cite{ryan2017self}. 
People who feel included develop a sense of belonging~\cite{sargent2002sense} that can significantly improve their psychological health~\cite{lambert2013belong,reker2013personal}.
For example, people with a strong sense of belonging have shown increased resilience~\cite{sargent2002sense}, and a strong sense of belonging in the workplace can mediate negative feelings like anxiety and depression~\cite{hartley2011examining,cockshaw2010link}. 
Furthermore, employees with a strong sense of belonging are more motivated to be productive and reach their fullest potential~\cite{isham2020wellbeing,walton2012mere,waller2020fostering}.
Similarly, perceptions of control and ownership positively contribute to employees' workplace satisfaction and job performance~\cite{Dyne2004PsychologicalOA,Eatough2014TheRO,pierce2004work}.}

\oldtext{Given the relationship between inclusion, control, and ownership and their implications for individual well-being in professional environments, it is crucial to explore how elements of workplace AI-mediated communication (AI-MC)~\cite{hancock-2020-ai} can impact these feelings, especially for users' from minoritized groups.}



\newtext{Large language models (LLMs) have the potential to transform workplace communication but can also exacerbate existing societal biases. LLMs are already seeing widespread use in assisting in everyday workplace tasks like writing emails and authoring documents. Recent scholarship has shown that workers can use LLMs to increase their productivity and produce higher-quality work~\cite{Zhang-2023-productivity}. 
Despite these potential benefits, LLMs also have drawbacks, including the potential to reproduce social biases~\cite{weidinger-etal-2022-risks,Bender-etal-2021-parrots}. 
For instance, when LLMs generate text about particular identity groups, they may display social biases in the content and style of the resulting text~\cite{wan2023kelly}, causing \textit{representational harm} through stereotyping and demeaning language~\cite{weidinger-etal-2022-risks,shelby2023}.}
\newtext{Beyond representational harms, we focus in this work on another category of harm: \textit{quality-of-service harm}~\cite{shelby2023,bird2020fairlearn}, in which systems unequally serve different groups, which may ``result in feelings of alienation, increased labor, and service or benefit loss'' for people of different identities~\cite{shelby2023}. 
For example, if an LLM produces text that is stylistically better aligned with one gender group over another, the result is a model that disparately benefits certain users over others. 
} 

\newtext{Such stylistic biases, whether real or perceived by users, can impact how individuals and groups of end-users feel about and interact with AI tools. When an LLM's text style better suits one group over another, such mismatch may impact users' feelings of inclusion when using that system.}
\newtext{Inclusion could affect other psychological factors like users' sense of the human experiences of control and ownership~\cite{ryan2017self} during the co-writing process.
For example, lower perceptions of inclusion may lead users to accept fewer AI suggestions; prior work has shown the relationship between accepting AI suggestions and feelings
of agency and control~\cite{mieczkowski2022ai}.}
\newtext{Given the potential importance of feelings of inclusion on people's use of AI tools, we set out to examine the relationship between writer-style bias and perceptions of inclusion, control, and ownership while co-writing with AI.}

In particular, we focus here on the potential impact of AI writing assistants that provide auto-complete suggestions as the user is typing text, for example, while writing an email. 
 In this context, we ask the following questions:
\begin{itemize}
  \item RQ1: Can the style of AI auto-complete suggestions impact people's sense of inclusion, control, and ownership over the text they write? 
  \item RQ2: How do these perceptions differ by gender?
\end{itemize}

More specifically, in this work, we investigate the impact of a specific communication style---assertiveness---on writers and its impact on minoritized gender groups. 
We build on a rich history examining gender disparities, particularly in the workplace where assertive language can be very consequential~\cite{amanatuallha-2010-negotiation,merchant-2012-gender-comm} and associated with better work outcomes~\cite{mazei2015meta,amanatuallha-2010-negotiation}.
While there are many cultural differences that guide expectations of gender presentation in the workplace, scholars in gender-based communication describe male talk in Western workplaces as being assertive, powerful, and authoritative~\cite{mullany-2007-gendered-discourse,merchant-2012-gender-comm,thimm-etal-2008-communicating}.
If writing assistants making suggestions in a more assertive style are not as well-aligned with the communication style of women and members of other minoritized gender groups, the discrepancy in style alignment could  
\newtext{cause an additional burden for women and gender minorities by increasing writing effort and task completion time.}
This paper addresses the research questions above using an online experiment where participants completed a hypothetical but common and consequential  workplace writing task: writing to a manager to ask for a promotion. 
Participants received auto-complete suggestions from one of two LLM-powered AI writing assistants, one assistant offering suggestions in an assertive style and the other in a hesitant style. 
We then asked participants about their feelings of inclusion, control, and ownership in the writing experience.


We found that the intervention---getting suggestions in an assertive or hesitant style from an AI assistant---did not impact participants' perceptions of inclusion. However, it did impact their perceptions of control and ownership. 
We did not find the assistant style treatments directly resulted in gender differences in our main measures. 
We do find evidence that gender plays some role:  minoritized genders' perceptions of control and ownership are more greatly impacted by their (likely pre-existing) feelings of inclusion than men's. In light of these findings, we offer a new conceptual model for understanding the impact of task- and user-style alignment on people's feelings of overall agency in AI-mediated environments.

\section{Background}

\oldtext{As large language models became more robust, with the ability to inspire ideas, revise text, and generate short stories, the field has shifted toward viewing other dimensions of the writing process~\cite{singh-etal-2022-creative,Bhat-cogpros-2023}. 
There is a growing interest in viewing writing assistants powered by large language models as co-authors and studying the interaction between users and these systems~\cite{lee-2022-coauthor,mirowski-screenplays-2023}.
Recent scholarship in this area has focused on understanding how writers evaluate and integrate the suggestions provided by these language models into their cognitive writing processes~\cite{Bhat-cogpros-2023}. 
Writers proactively engage with system-generated content, making deliberate efforts to integrate it into their writing~\cite{singh-etal-2022-creative}.
However, there is evidence that perceptions of ownership can differ.
Inexperienced writers perceive less ownership than experienced writers during the co-writing process~\cite{mieczkowski2022ai}.
Here, we expand the understanding of how these systems can best support and enhance the writing process by demonstrating the importance of inclusion when interacting with an AI writing assistant.} 

\oldtext{As large language models like GPT~\cite{vaswani2023attention,bommasani2022opportunities} are increasingly integrated into our society, 
there is growing concern among AI ethicists regarding their potential harms, including in the context of auto-complete suggestions. 
Concerns have been raised regarding the risks of misinformation, privacy violations, socioeconomic harms, and, most salient to our work here, representational harms~\cite{weidinger-etal-2022-risks}.  
Large language models are trained on large corpuses of data which often reflect historic and systemic injustices. 
For example, a language model is more likely to produce the phrase ``he is a doctor'' than ``she is a doctor'', assuming the subject's gender or ethnic identity~\cite{weidinger2021ethical}. 
As a result, these models may produce demeaning language and stereotypes which can be especially harmful for some with intersecting identities. 
Specifically in our context, biases in auto-complete suggestions have been shown to shift the use of language and influence content~\cite{hohenstein2023artificial,poddar2023ai,jakesch-etal-2023-cowriting}. 
AI suggestions even shift people's opinions~\cite{jakesch-etal-2023-cowriting}.
Extending this body of work, here we show how AI biases may impact the perceptions of people using these technologies.}

\newtext{As LLMs become more robust with the ability to inspire ideas, revise text, and generate short stories, the writing assistants powered by large language models are increasingly seen as co-writers. There is a growing body of work studying the interaction between users and these co-writing systems~\cite{lee-2022-coauthor,mirowski-screenplays-2023}. 
Recent scholarship in this area has focused on understanding how writers evaluate and integrate the suggestions provided by LLMs into their cognitive writing processes~\cite{Bhat-cogpros-2023}, and how writers proactively engage with system-generated content~\cite{singh-etal-2022-creative}.}

\newtext{Along with their potential benefits, however, these models carry potential risks---risks that may not be equally distributed across different user groups. Prior work has identified several categories of harms including social systems harms,
allocative harms, representational harms, quality of service harms, and interpersonal harms~\cite{shelby2023,weidinger-etal-2022-risks}.
Prior work has pointed to the existence of such harms in the context of auto-complete suggestions.
For example, societal biases in the suggestions can cause information harms (a subset of social systems harms) by influencing what content is written ~\cite{poddar2023ai,jakesch-etal-2023-cowriting}, and even shifting writers' attitudes, potentially without their awareness~\cite{jakesch-etal-2023-cowriting,kidd-birhane2023}.
While we do not know of work showing representational harms in auto-complete environments, given the potential for content influence~\cite{jakesch-etal-2023-cowriting,kidd-birhane2023} and the known representational biases that exist in LLM text generation~\cite{shelby2023,weidinger-etal-2022-risks}, the issue is likely to persist in the auto-complete context as well.}  


\newtext{Quality-of-service harms are another concern identified by researchers.
These harms reflect developer's choices and result in unintended performance disparities based on identity ~\cite{shelby2023,bird2020fairlearn}.
For example, prior work has shown that audio speech recognition (ASR) systems have significantly higher error rates for Black speakers than White speakers~\cite{koenecke2020asr}. These error rates can have psychological and behavioral impacts on Black users such as feelings of frustration and disappointment; they can cause users to have to modify their linguistic patterns~\cite{mengesha2021don}. Such quality-of-service harms can lead to alienation and result in greater labor burdens and loss of benefits for marginalized communities~\cite{shelby2023}.
In this short paper, we extend research on this topic to consider writer-style biases in auto-complete suggestions and their potential to cause  differential impacts on  inclusion, control, and ownership for different  social groups, with a specific focus on gender.}


\section{Methods}

We addressed our research questions using an online experiment where participants completed a common workplace writing task. 
We asked participants to write a hypothetical email to their manager asking for a promotion. 
We chose the promotion task since it is a situation that requires high assertiveness in context in which minoritized genders face several challenges~\cite{amanatuallha-2010-negotiation,Amanatullah2013AskAY,mazei2015meta}. 
The experiment had three conditions. 
Participants in two of the treatments received auto-complete suggestions from a language model. 
In one of these treatments, the model made \textit{assertive} auto-complete suggestions, and in the other, it made \textit{hesitant} auto-complete suggestions. 
A control group of participants wrote without the assistance of any model.

After the writing task, participants answered a demographic questionnaire. 
Participants in the AI conditions answered an additional post-task questionnaire regarding their perceptions of inclusion, ownership, and agency while writing with the AI assistant. 
The questions and 5-point Likert scale answers were adapted from previous work~\cite{mieczkowski2022ai, jakesch-etal-2023-cowriting}. 
The research design was approved by Cornell's IRB. 
The design was preregistered and is available on OSF\footnote{\url{https://osf.io/kwbev?view_only=c5639dcc6292482b8bd79f2086b38d24}}.

\subsection{Writing App}

\begin{figure*}[ht]
\includegraphics[width=0.9\textwidth]{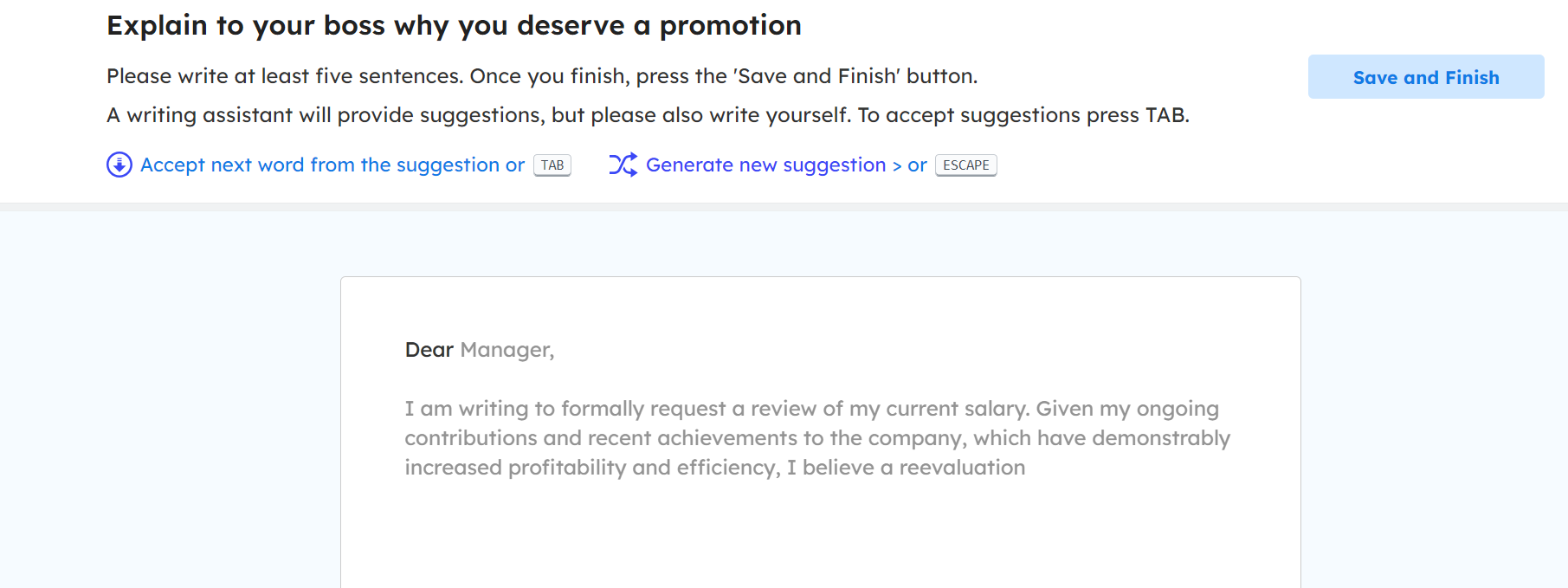}
\caption{\textbf{Screenshot of the writing task.} Instructions are given in the panel at the top. Participants can hit `tab' to accept the suggestion or `esc' to generate a new suggestion. 
Written text is in black and suggestions appear in gray text.
}
\label{fig:interface}
\end{figure*}

We used a custom experimental platform combining a rich-text editor and a writing assistant, as seen in Figure~\ref{fig:interface}. 
The platform code was adapted from prior experimental work on auto-complete systems~\cite{jakesch-etal-2023-cowriting}. 
The suggestions work such that when the participant pauses typing, the system displays suggestions for completion that are 20~words or less.
The user can then choose to accept the suggestion by pressing \textit{tab} or \textit{right arrow key,} or they can ignore the suggestion by continuing to type.  
Pressing \textit{tab} accepts the subsequent word in the suggestion, and a user can press multiple times to accept the full suggestion.  

To generate the interactive text suggestions for the experiment, we used GPT-4 via an API call with prompts carefully designed to display assertive or hesitant text. 
To achieve that goal, we inserted the prompt ``In a [self-assured/hesitant] manner ask for a raise'' before the participants’ written text. After the entered text, we added the affirming instruction ``add one sentence continuing the email in a [self-assured/hesitant] manner asking for a raise''. 
This way, the suggestions maintained the desired style regardless of the text already entered by the participant. 

\subsection{Participant Recruitment}

We recruited 750 participants from a gender-balanced sample of English-speaking US-based adults on the crowdsourcing service Prolific. The sample size was calculated using a two-way ANOVA with a power of 80\% based on a pre-test survey. 
\newtext{To} ensure response quality, we required participants to have approval ratings greater than 95\% on Prolific. 
We excluded participants who did not state their gender and manually verified that each writing sample was faithful to the writing task. 
In the end, we had usable data from 738 participants. 

Our experiment considers two distinct gender categories: men, as category that is usually associated with greater societal influence and dominance, and minoritized genders who typically encounter societal disparities.
While describing our results, we will refer to \textbf{W}omen, \textbf{N}on-\textbf{B}inary or third gender, and ``preferred to describe'' (also known as \textbf{S}elf-\textbf{D}escribed) as \textbf{WNBSD} participants. 
Of the 738 participants whose data is used, 48.9\% identified as men and 51.1\% as WNBSD: 47.7\% identified as women, 3\% identified as non-binary or third gender, and 0.4\% preferred to describe their gender. 

The majority of the participants were between 25-34 years old (56\%), identified as white (61.9\%), received a bachelor’s degree (38.2\%), reported working a full-time job (59.4\%), and reported working entirely in person (40.6\%). 

\subsection{Measures}
\label{sec:measures}

We used the following measures to examine how participants interacted with the AI assistant and their perceptions of the writing task with the AI assistant.

\textbf{AI reliance}. 
Prior work on how users interact with AI writing assistants uses the measure of mutuality to describe the level of \textit{interaction} the writer has with the AI assistant~\cite{mieczkowski2022ai,lee-2022-coauthor}. 
\newtext{However, in this work, we are interested in the final text output, not the event-based interactions.}
\newtext{We introduce a new measure, \textit{AI reliance}, as we believe this measure captures the construct more intuitively. AI reliance is the fraction of AI-written characters in the entire text. The measure uses a score from~0 to~1, where~1 means the assistant wrote the entire text (reflecting complete reliance on the assistant) and 0 means the human wrote the entire text (no reliance).}


\textbf{Inclusion}.
We draw on the concept of self-extension, which explains how objects become integral to one's self-concept~\cite{belk2013extended}, to develop the notion of inclusion in AI-mediated communication.
We focus on identity self-extension where the technology is a reflection of the user~\cite{ross2021explicating,mieczkowski2022ai}.
This literature suggests that if a user feels the technology reflects themselves, they will feel included. 
We therefore measure inclusion with the following statements: \textit{the writing assistant was made for people like me}, and \textit{the writing assistant sounded like I would write myself}.
Response options ranged from \textit{strongly disagree} (1) to \textit{strongly agree} (5).

\textbf{Control}.
Prior work has shown that feeling in control is not limited to one action or outcome. 
Feeling in control can encompass engaging in an action, control over the outcome, or both~\cite{singh-etal-2022-creative,mieczkowski2022ai}.
For example, some users may feel a lack of control since they cannot choose which suggestions they are seeing (control during the writing process). 
However, the fact that they can select which suggestions are incorporated into the message can impact their sense of control over the process \textit{and} the final version~\cite{Bhat-cogpros-2023}. 
We therefore measure control by asking participants \textit{how much control did you feel over the process of writing the message} and \textit{how much control did you feel over the final version of the message}~\cite{mieczkowski2022ai}. 
Response options ranged from \textit{not at all} (1) to \textit{a great deal} (5).

\textbf{Ownership}.
One aspect of ownership involves "mineness"---the idea that a specific target belongs to an individual. 
The target can include physical objects but can also encompass intangible objects such as words and thoughts.
In our experiment, we consider two targets---the message itself and the message style.
Therefore, we also investigate ownership over the writing content and style. 
Our questions included \textit{to what extent do you feel like the message you wrote is yours} and \textit{thinking back on the message writing activity, how much did the message sound like you}~\cite{mieczkowski2022ai}.
The response options were the same as those for the control questions.


\section{Results}
\label{sec:results}
Our results expose the impact of writing style on perceptions of inclusion, control, ownership, and agency and hint at potential differences between genders in that respect.
Our primary, preregistered analysis finds some significant effects of AI writing style on perceptions of control and ownership but not on inclusion.
\newtext{A follow-up, exploratory
analysis suggests that perceptions of inclusion mediate the loss of agency as AI reliance increases.}
Additionally, the analysis suggests that heightened perceptions of inclusion lead to increased agency, particularly among minoritized genders.

We first provide an overview of the effects of using the AI auto-complete suggestions on the writing task.
We examined three components of the writing task: the length of text written by participants, their writing time, and their degree of AI reliance. 
We performed a two-way ANOVA to compare the means between the three treatment groups (self-assured AI writing style, hesitant AI writing style, and control) and two gender groups (men and WNBSD participants).
Overall, participants who wrote with the assistant wrote longer emails (M=160.18, SD=82.16) and spent less time on the task (M=303.82, SD=218.53) than participants in the control group (email length M=95.53, SD=35.64; time on task M=363.66, SD=292.68).
The differences are statistically significant (length: ANOVA F=67.86, \textit{p}<0.001; time: ANOVA F=6.35, \textit{p}=0.002).
Post-hoc tests did not observe any differences between the two AI conditions in email length and time on task. 
In contrast, we \textit{did} observe significant impacts of AI writing style on AI reliance: participants who wrote with the self-assured writing style model relied on the model more (M=0.55, SD=0.34) than participants who wrote with the hesitant writing style model (M=0.45, SD=0.35).
These differences were also significant (t=3.16, \textit{p}=0.002).
We did not observe gender effects or interaction effects between gender and the AI treatment on any of these metrics.

\subsection{Inclusion}
\begin{figure}[h]
\includegraphics[width=\columnwidth]{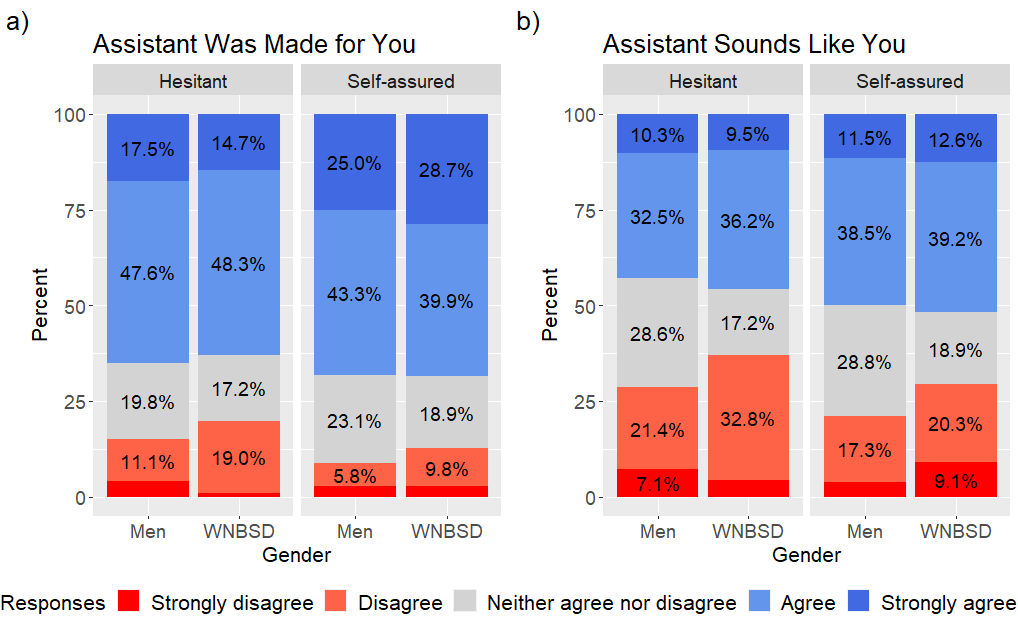}
\caption{\textbf{Participants’ assessment of inclusion.} \newtext{Participants are more likely to say the assistant was made for them than sounds like them. Responses less than 5\% are not labeled on the figure.}\oldtext{Responses less than 5\% are not labeled on the figure.}
}
\label{fig:inclusion}
\end{figure}


We examine the impact of AI writing style on our key measures, beginning with perceptions of inclusion.
\newtext{Figure~\ref{fig:inclusion} shows the frequency of different survey responses for each gender condition (different columns) for the inclusion statements: \textit{the writing assistant was made for people like me} (``Assistant Was Made for You'' on the left), and \textit{the writing assistant sounded like I would write myself} (``Assistant Sounds Like You'' on the right).}
\newtext{
In the condition with the hesitant AI writing style (left side of Figure~\ref{fig:inclusion}a),
65.1\% of men (first column) and 63\% of WNBSD participants (second column) either agreed or strongly agreed (blue and dark blue bars in the column) that the writing assistant was made for people like them. 
As the figure also shows, in the self-assured AI writing style group (right side of Figure~\ref{fig:inclusion}a), 
68.3\% of the men and 68.6\% of WNBSD participants either agreed or strongly agreed with this statement.
In general, participants more readily agreed with the ``made for you'' statement (Figure~\ref{fig:inclusion}a) compared to the ``sounds like you'' statement (Figure~\ref{fig:inclusion}b).} 

\oldtext{Figure~\ref{fig:inclusion} shows, for each gender condition (different bars) the frequency of different survey responses for the inclusion statements: 
\textit{the writing assistant was made for people like me} (``Assistant Was Made for You'' on the left), and \textit{the writing assistant sounded like I would write myself} (``Assistant Sounds Like You'' on the right).}

\oldtext{Figure~\ref{fig:inclusion}a (top-left) shows that in the condition with the self-assured AI writing style, 68.3\% of the men and 68.5\% of the WNBSD participants either agreed or strongly agreed (the blue and dark blue shades on the right of the bars) that the writing assistant was made for people like them. 
As the figure also shows, in the hesitant AI writing style group, (the bottom-left Figure~\ref{fig:inclusion}c), 
65.1\% of men and 62.9\% of WNBSD participants either agreed or strongly agreed with this statement. 
}


\oldtext{Figure~\ref{fig:inclusion}b (top-right) shows the participants who wrote with the self-assured writing style model; of the participants who received this treatment, 50\% of the men and 51.7\% of the WNBSD participants either agreed or strongly agreed that the assistant sounded like how they would write themselves.
In the condition with the hesitant AI writing style, (bottom-right Figure~\ref{fig:inclusion}d), 42.9\% of the men and 45.7\% of the WNBSD participants either agreed or strongly agreed. 
A two-way ANOVA analysis revealed no significant effects for gender, AI treatment group, or the interaction between gender and group across all inclusion questions.}


\newtext{
In Figure~\ref{fig:inclusion}b, of the participants who received the hesitant treatment (left), 42.8\% of the men and 45.7\% of the WNBSD participants either agreed or strongly agreed that the assistant sounded like how they would write themselves. 
In the condition with the self-assured AI writing style, on the right side of Figure~\ref{fig:inclusion}b, 50\% of the men and 51.8\% of the WNBSD participants either agreed or strongly agreed. A two-way ANOVA analysis revealed no significant effects for gender, AI treatment group, or the interaction between gender and treatment group across both inclusion questions shown in Figure~\ref{fig:inclusion}.}

\subsection{Control}
\begin{figure}[h]
\includegraphics[width=\columnwidth]{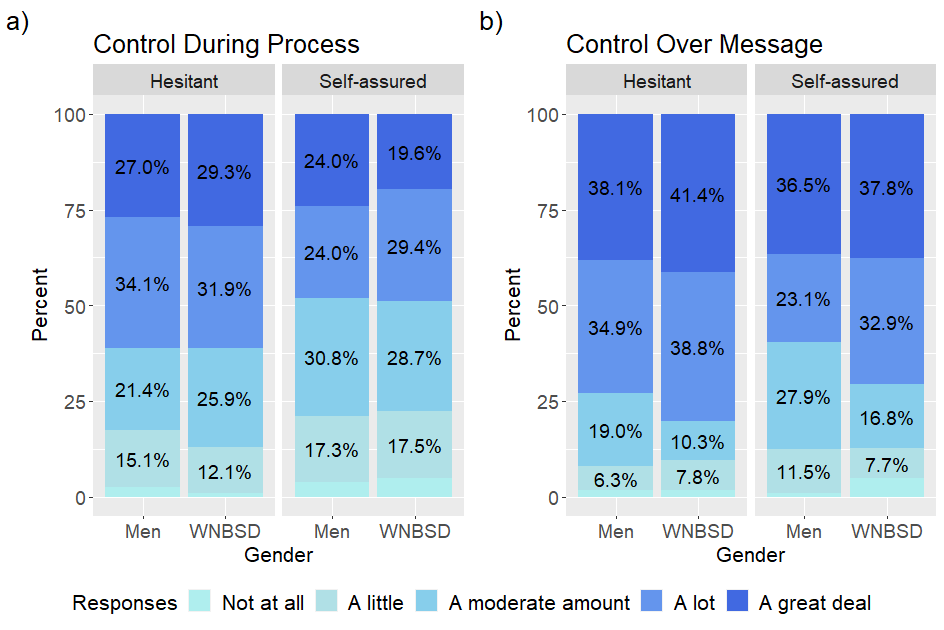}
\caption{\textbf{Participants assessment of control.}  Participants assisted by the hesitant writing style model are more likely to feel greater control over the final version of the message and the writing process than participants assisted by the self-assured writing style model. Responses less than 5\% are not labeled on the figure. 
}
\label{fig:control}
\end{figure}



\oldtext{Figure~\ref{fig:control} illustrates, for each gender condition \oldtext{(different rows)} \newtext{(different columns)} the frequency of different survey responses for the following questions: \textit{how much control did you feel over the process of writing the message} (``Control During Process'' on the left), and \textit{how much control did you feel over the final version of the message} (``Control Over Message'' on the right).}

\newtext{Figure~\ref{fig:control} shows the frequency of different survey responses for each gender (different columns) for the following questions: \textit{how much control did you feel over the process of writing the message} (``Control During Process'' on the left), and \textit{how much control did you feel over the final version of the message} (``Control Over Message'' on the right).}
\newtext{In the condition with the hesitant AI writing style suggestions (left side of Figure~\ref{fig:control}a), 61.1\% of the men (first column) and 61.2\%  of WNBSD participants (second column) reported feeling a lot or great deal of control (the darker blue and darkest blue bars in the columns) during the writing process.
In the condition with the self-assured AI writing style (right side of Figure~\ref{fig:control}a), 48\% of men and  49\% of the WNBSD participants expressed a lot or a great deal of control during the writing process.}
A two-way ANOVA revealed that the difference between the AI treatment groups was significant (\textit{p}=0.004),
meaning that participants in the hesitant condition reported more control over the process. 
There were no significant differences for gender or the interaction between gender and the AI treatment groups for this question.
\oldtext{Figure~\ref{fig:control}a (top-left) shows in the condition with the self-assured AI writing style suggestions,
during the writing process, 48\% of men and  49\% of WNBSD participants reported feeling a lot or a great deal of control (indicated by the darker blue and darkest blue within the columns).
In the condition with the hesitant AI writing style, seen in Figure~\ref{fig:control}c (bottom-left), 61.1\% of the men and 61.2\% of the WNBSD participants expressed a lot or a great deal of control during the writing process.}

\oldtext{In the condition with the self-assured AI writing style, (Figure~\ref{fig:control}b, top-right), 59.6\% of men and 70.6\% of WNBSD participants reported a lot or a great deal of control during the writing process. 
Figure~\ref{fig:control}d (bottom-right) shows 73\% of men and 80.2\% of WNBSD participants who wrote with a hesitant writing style model reported a lot or a great deal of control over the final version of the message.}

\newtext{Looking at Figure~\ref{fig:control}b and control over the final version, in the condition with the hesitant AI writing style (left), 73\% of men and 80.2\% of WNBSD participants reported a lot or a great deal of control over the final version of the message. 
The right side of Figure~\ref{fig:control}b shows 59.6\% of men and 70.7\% of WNBSD participants who wrote with a self-assured writing style model reported a lot or a great deal of control over the final version of the message.}
For this question as well, a two-way ANOVA revealed that the difference between the AI treatment groups was significant (\textit{p}=0.046),
with participants in the hesitant AI writing style condition reporting greater control over the final version of the message. 
At the same time, there were no significant differences in relation to gender or interactions between gender and the AI treatment group for this question.
In both treatment conditions, participants generally reported higher levels of control over the final version of the message (Figure~\ref{fig:control}b) than the during the writing process (Figure~\ref{fig:control}a).

\subsection{Ownership}

\begin{figure}[h]
\includegraphics[width=\columnwidth]{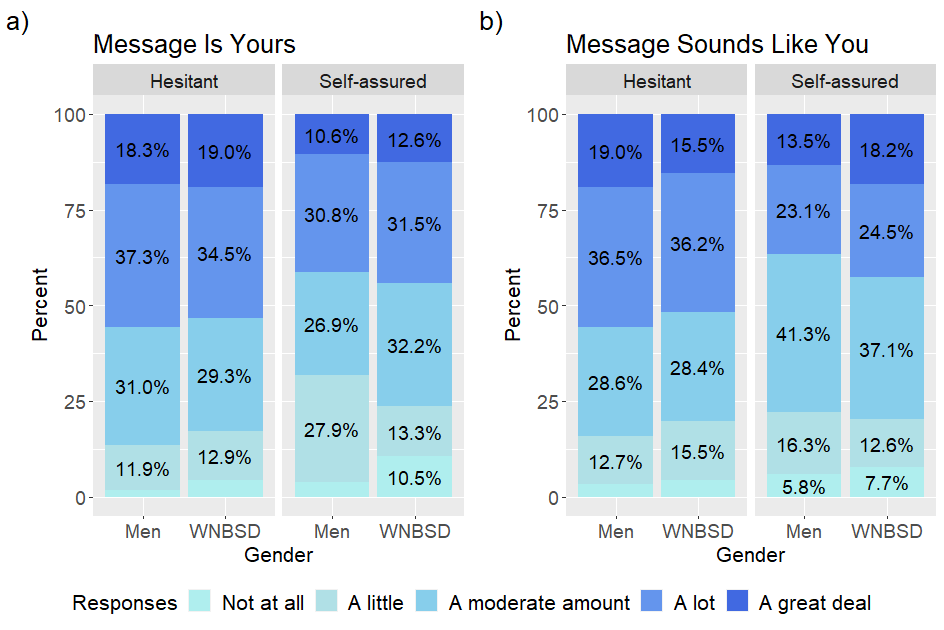}
\caption{\textbf{Participants assessment of ownership.} Participants assisted by a hesitant model are more likely to say that they wrote the message and the message sounds like them. Responses less than 5\% are not labeled on the figure.}
\label{fig:ownership}
\end{figure}



Figure~\ref{fig:ownership} illustrates the frequency of different survey responses for each gender condition\oldtext{(rows)} \newtext{(columns)} for the following questions: \textit{to what extent do you feel like the message you wrote is yours} (``Message Is Yours''), and \textit{thinking back on the message writing activity, how much did the message sound like you} (``Message Sounds Like You'').
\newtext{In the hesitant AI writing style condition (left side of Figure~\ref{fig:ownership}a),  55.6\% of the men and  53.5\%  of WNBSD participants expressed a lot or a great deal of ownership over the message (the dark blue and darkest blue bars in the column).
In the self-assured AI writing style condition (right side of Figure~\ref{fig:ownership}a), participants experienced a higher level of ownership, with 41.4\% of men and 44.1\% of the WNBSD participants expressing a similar degree of ownership over the message.}
A two-way ANOVA revealed that this difference between the AI treatment groups was significant 
(\textit{p}<0.001). 
There was no significant difference for gender or the interaction between gender and the AI treatment groups for this question.

\oldtext{In the self-assured AI writing style condition, (Figure~\ref{fig:ownership}a, top-left), 41.4\% of men and 44.1\% of WNBSD participants expressed a lot or a great deal of ownership over the message (as seen by the dark blue and darkest blue within the columns).
In the hesitant AI writing style condition (Figure~\ref{fig:ownership}c, bottom-left), participants experienced a higher level of ownership with 55.6\% of the men and  53.5\% of the WNBSD participants expressing a similar degree of ownership over the message.}

\oldtext{Reflecting on the style of the message, in the self-assured AI writing style condition figures~\ref{fig:ownership}b and d on the right show similar trends: participants in the hesitant AI writing style condition (bottom) were more likely to say the final message reflected their style.} 
\newtext{Reflecting on the style of the message, Figure~\ref{fig:ownership}b shows a similar trend: participants in the hesitant AI writing style condition (left) were more likely to say the final message reflected their style.}
Once again, a two-way ANOVA revealed that the difference between the AI treatment groups was significant \oldtext{(p<0.05)}(\textit{p}=0.032).
There was no significant difference for gender or the interaction between gender and the AI treatment groups and gender for this question.

Overall, our results show AI writing style impacts control and ownership.
Participants who wrote with the hesitant model expressed greater control over the writing process and the final message and felt the message sounded more like them than participants who wrote with the self-assured model.
However, our analysis did not reveal an impact of the AI writing style on perceived inclusion.
Additionally, we did not see any effects of gender or interaction between gender and treatment group across inclusion, control, and ownership questions.
We further explore these factors in the analysis below. 

\subsection{Exploratory Analysis: Inclusion and Agency}
\label{sec:exploratory}

To better understand the relationship between the different measures and the participants' interaction with the AI model, we perform an exploratory analysis that brings together multiple factors using a linear regression model. 
To perform this analysis, we combined various measures into simplified high-level constructs. 
First, we use a compound measure of agency as our dependent variable.
The concept of agency has been defined in several different ways but broadly captures the idea of ``the capacity to alter a situation''~\cite{neff2018agency,cesafsky2019calibrating}. 
The variables that reflect that in our study were control and ownership. 
While distinct, the literature shows these concepts are closely related:
whether a person has control greatly influences their perceived ownership~\cite{mieczkowski2022ai,braun2018senses}.
An analysis of our measures confirmed the control and ownership questions were indeed highly correlated in our data.
We thus created a score for agency by performing a row-wise average across the control and ownership questions.

The independent variables in the model consisted of a simplified inclusion variable, the AI reliance score, the AI writing style, and demographic variables. 
To simplify the model and analysis, we combined the two highly correlated inclusion measures into one variable by performing a row-wise average across the two inclusion questions 
\newtext{(Table~\ref{table:correlation_questions} in the appendix shows the correlation of the inclusion, control, and ownership measures).}
The model also included the measure of AI reliance as defined above, namely how much the participants accepted the AI suggestions in the writing process. 
Finally, we include in our model the AI writing style of the model that participants wrote with (the AI treatment), as well as more detailed demographic variables: age, gender, race, and income. 
The race and gender features were operationalized as binary variables where 1 reflected the privileged group and 0 reflected the minoritized group (e.g., for gender, 1=men, 0=WNBSD).
We also mapped the AI writing style as a binary variable (0=hesitant model, 1=self-assured model).
We operationalized the age and income demographic features as ordinal variables. 
Note again that we consider this analysis exploratory, as it was not part of the preregistered analysis presented above that focused on the main effects.

\begin{table}[t] \centering 
  \caption{OLS linear regression analysis predicting perceived agency based on demographic features, AI writing style, AI reliance, and perceived inclusion. The constant corresponds to the baseline perceived agency.} 
  \label{table:regression} 
  \footnotesize
\begin{tabular}{@{\extracolsep{5pt}}lc} 
\\[-1.8ex]\hline 
\hline \\[-1.8ex] 
 & \multicolumn{1}{c}{\textit{Dependent variable:}} \\ 
\cline{2-2} 
\\[-1.8ex] & Agency \\ 
\hline \\[-1.8ex] 
 Age & 0.024 \\ 
  & (0.020) \\ 
 Race & 0.013 \\ 
  & (0.070) \\ 
 Income & 0.023 \\ 
  & (0.034) \\ 
 Gender & 0.571$^{*}$ \\ 
  & (0.261) \\ 
 `AI Reliance` & $-$3.451$^{***}$ \\ 
  & (0.367) \\ 
 Inclusion & 0.136$^{*}$ \\ 
  & (0.066) \\ 
 `AI Writing Style` & $-$0.244$^{**}$ \\ 
  & (0.090) \\ 
 Gender:`AI Reliance` & 0.251 \\ 
  & (0.191) \\ 
 `AI Reliance`:Inclusion & 0.666$^{***}$ \\ 
  & (0.098) \\ 
 Gender:Inclusion & $-$0.204$^{**}$ \\ 
  & (0.071) \\ 
 Gender:`AI Writing Style` & 0.050 \\ 
  & (0.133) \\ 
 Constant & 3.661$^{***}$ \\ 
  & (0.248) \\ 
\hline \\[-1.8ex] 
Observations & 489 \\ 
R$^{2}$ & 0.354 \\ 
Adjusted R$^{2}$ & 0.339 \\ 
Residual Std. Error & 0.715 (df = 477) \\ 
F Statistic & 23.747$^{***}$ (df = 11; 477) \\ 
\hline 
\hline \\[-1.8ex] 
\textit{Note:}  & \multicolumn{1}{r}{$^{*}$p$<$0.05; $^{**}$p$<$0.01; $^{***}$p$<$0.001} \\ 
\end{tabular} 
\end{table}

Table~\ref{table:regression} shows the parameter estimates of a linear regression model predicting perceived agency.
We tested several versions of the model that accounted for additional interactions (like AI writing style and AI reliance), and we present the model with the best fit. 
The model shows a main effect of AI writing style (Table ~\ref{table:regression}, line 7) on agency.
The negative value of the parameter signifies that writing with a self-assured model contributes to a loss of agency.
We also see main effects of gender (line 4), showing men contributed to higher levels of agency. 
Perceptions of inclusion (line 6) also positively impacted agency, and AI reliance (line 5) had a negative impact.
These effects were qualified by interaction effects between AI reliance and perceived inclusion (line 9) and gender and perceived inclusion (line 10), which we explore next.


\begin{figure*}[t]
    \centering
    \begin{subfigure}[t]{0.48\textwidth}
        \centering
        \includegraphics[width=\textwidth]
        {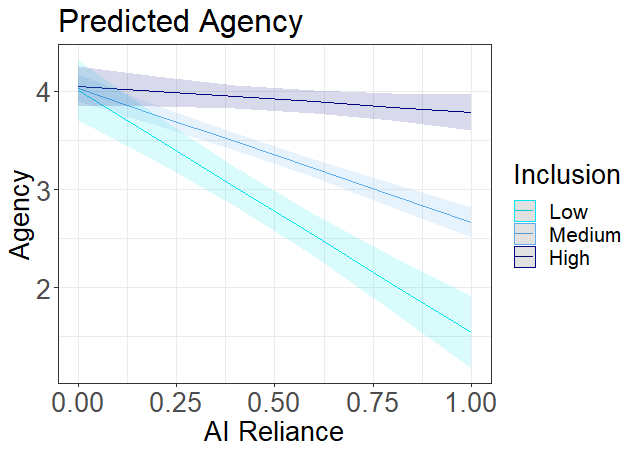}
        \caption{Participants who feel higher level of inclusion by the writing assistant maintain agency even when AI reliance increases}
        \label{fig:interaction_ai}
    \end{subfigure}%
    \begin{subfigure}[t]{0.48\textwidth}
        \centering
        \includegraphics[width=\textwidth]
        {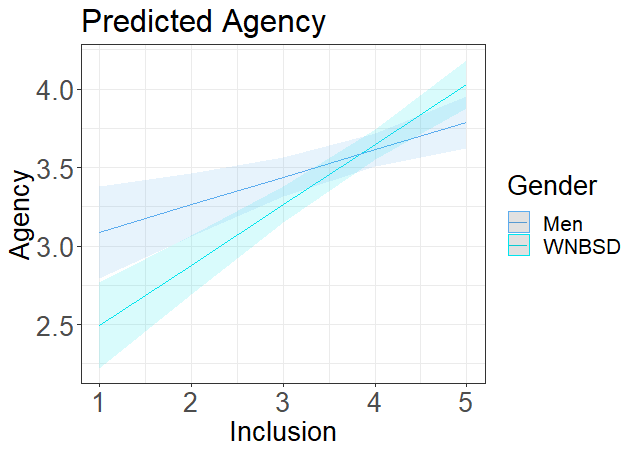} 
        \caption{Gender minorities have a stronger sense of inclusion}
        \label{fig:interaction_inc}
    \end{subfigure}
    \caption{\textbf{Interaction Effects in the Linear Regression}. Figure~\ref{fig:interaction}a depicts the interaction between Inclusion and AI Reliance. Figure~\ref{fig:interaction}b depicts the interaction between Gender and Inclusion.}
    \label{fig:interaction}
\end{figure*}

We explore the interaction effects to better understand their impact on perceived agency.
Figure~\ref{fig:interaction_ai} depicts the interaction between AI reliance and perceived inclusion. 
AI reliance is the x-axis, and perceived agency is the y-axis.
The graph compares the groups that showed high (dark blue), medium (blue), and low (light blue) levels of inclusion. 
Participants with a high level of inclusion, as indicated by the dark blue line in the figure, maintain their sense of agency even with increasing reliance on the writing assistant. 
Participants with a low (and to a lesser degree, medium) sense of inclusion lose their sense of agency rapidly when they rely more on AI.

Figure~\ref{fig:interaction_inc} explores how the interaction between inclusion (x-axis) and gender impacts agency. The graph compares men (blue line) and minoritized genders (WNBSD, light blue line).
While inclusion and agency have a direct relationship for all gender identities, the slope for WNBSD participants is greater than the slope for men, indicating that perceived inclusion is more strongly tied to perceived agency for minoritized genders.

\section{Discussion}
Our findings enhance the current models of ``writing with AI'' by introducing new factors and furthering our understanding of their relationships. The findings also hint at the role that gender may play in such a context, though this line of investigation requires additional substantiation. 
Overall, the findings offer implications for research and design of co-writing interactions with AI.  

By introducing the construct of \textit{inclusion}, our work extends the previous research on AI-MC, which has examined perceptions of agency while interacting with an AI writing assistant~\cite{mieczkowski2022ai, mieczkowski2021ai}. 
We measure the concept of inclusion after the co-writing process by asking participants if they could envision themselves in the model 
and whether the model's writing style matched their own. 
We show that this concept contributes to the perception of agency, a combined metric reflecting ownership and control over the writing process and the final product. 

\begin{figure}[t]
\includegraphics[width=\columnwidth]{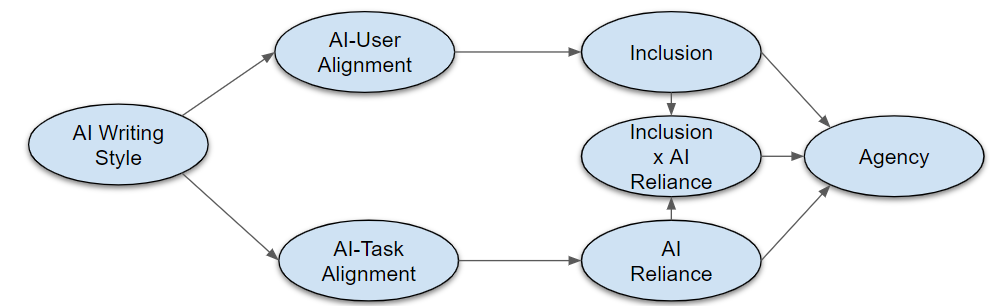}
\caption{\textbf{Conceptual Model of Agency:} A proposed model of the constructs and measured variables that contribute to feelings of agency.}
\label{fig:model}
\end{figure}

Informed by these findings, Figure~\ref{fig:model} presents a new conceptual model of the factors contributing to agency in AI co-writing environments. 
This model is tentative: not all of these concepts were measured, let alone tested, in our work here. 
The model suggests that there are two paths for feeling agency in the co-writing process and its outcome. 
On the top, the AI writing style affects AI-user alignment---which we define as the similarity between the AI writing style and the user's own preferences that may be derived from their identity. 
High AI-user alignment leads to increased perceived inclusion, positively contributing to perceived agency.
In our research, we did not measure AI-user alignment. Instead, we built on the literature suggesting that minoritized genders are less likely to be aligned with assertive AI suggestions~\cite{merchant-2012-gender-comm,Herring-etal-2004-authenticity}---which was not the case in this study. 
At the bottom of the model, the AI writing style affects AI-task alignment, which we define as the perceived usefulness of the AI assistant for the task (e.g., writing to ask for a promotion with an assertive AI). 
If there is high AI-task alignment, the user will rely on the model to complete the task. 
However, such reliance, as we have shown, can \textit{negatively} contribute to the user's perceived agency. 
The conceptual model includes an interaction between AI reliance and inclusion, as suggested by our exploratory results. 

Our model raises several open questions that have implications for designers.
The most salient question is the potential competition between AI-User alignment and AI-Task alignment. 
Or, how can developers strike a balance between models that are inclusive of various users but are also effective in completing the task?
We note that even if the models align with both the user and task, there is remaining conflict: the AI-User alignment will ultimately contribute to agency, but the AI alignment with the task---which will increase reliance---will also \textit{reduce} agency.
This finding complicates the question of how developers can build models that help the user accomplish the task while also supporting the user's agency, given the tension between those.
\newtext{Further work is needed to understand the trade-off between agency and task performance in real-world applications.}

The idea of inclusion via AI-User alignment captures our hypothesis that gender identity will play a role in perceived inclusion and agency. 
We hypothesized that minoritized genders would feel more included in the hesitant AI writing style condition, at least compared to men. 
However, our main findings did not show such gender effects. 
Although existing literature highlights language use differs by gender in negotiation contexts, it often relies on experimental methods (such as laboratory studies) that encompass factors beyond the scope of our experiment like job status, vocal pitch, or physical presence~\cite{amanatuallha-2010-negotiation,mazei2015meta}.
Many of the laboratory studies show it is the small details, like the tone of voice or the wording, which makes gender differences salient~\cite{thimm-etal-2008-communicating}.
The absence of physical cues, as shown in computer-mediated negotiations, reduces gender-based communication bias~\cite{stuhlmacher1999gender}.


While gender differences did not account for our expected main effect, our exploratory analysis also suggests there are still, potentially, \textit{some} differences between genders.
A noteworthy finding in this analysis was the significant interaction between gender and inclusion in relation to perceived agency. 
The interaction suggests that for minoritized genders, perceptions of inclusion have a greater effect on the perception of agency compared to men. 
If they hold, these findings lend support to our initial hypothesis: that people of different genders may react differently to writing assistants of different styles. 
This finding suggests that it may be useful to devise, develop, and test language models that also optimize on generating perceptions of inclusion by people of different demographic backgrounds. 

\newtext{
Our work explored the impact of AI auto-complete suggestions on feelings of inclusion, control, and ownership over the written text and considered the implications for users of various backgrounds. 
The work has a number of limitations.
First, in asking people to write as if asking for a promotion, our study focused on an important and realistic workplace task. 
However, other workplace writing tasks should also be studied to generalize the results.
Furthermore, we did not measure individual differences in assertiveness, which could have provided additional insight on the effect of the treatment. 
Pre-task measurement of such attitudes, though, could have introduced bias to the experimental task. 
We focus on English-speaking US residents.
Additional work is needed to understand how users from various linguistic backgrounds and cultures interact with such AI agents. 
Despite these limitations, this short paper showed some of the first evidence that minoritized genders’ perceptions of control and ownership could vary in some ways from men’s as a response to writer-style biases. 
We offer a new conceptual model for understanding the impact of these factors on people’s feelings of overall agency in AI-mediated environments.}

\begin{acks}
This material is based upon work supported by the National Science Foundation under Grant No. CHS 1901151/1901329.
\end{acks}

\bibliographystyle{ACM-Reference-Format}
\bibliography{aimc-gender-bibliography}


\begin{thebibliography}{31}


\ifx \showCODEN    \undefined \def \showCODEN     #1{\unskip}     \fi
\ifx \showDOI      \undefined \def \showDOI       #1{#1}\fi
\ifx \showISBNx    \undefined \def \showISBNx     #1{\unskip}     \fi
\ifx \showISBNxiii \undefined \def \showISBNxiii  #1{\unskip}     \fi
\ifx \showISSN     \undefined \def \showISSN      #1{\unskip}     \fi
\ifx \showLCCN     \undefined \def \showLCCN      #1{\unskip}     \fi
\ifx \shownote     \undefined \def \shownote      #1{#1}          \fi
\ifx \showarticletitle \undefined \def \showarticletitle #1{#1}   \fi
\ifx \showURL      \undefined \def \showURL       {\relax}        \fi
\providecommand\bibfield[2]{#2}
\providecommand\bibinfo[2]{#2}
\providecommand\natexlab[1]{#1}
\providecommand\showeprint[2][]{arXiv:#2}

\bibitem[Amanatullah and Morris(2010)]%
        {amanatuallha-2010-negotiation}
\bibfield{author}{\bibinfo{person}{Emily Amanatullah} {and}
  \bibinfo{person}{Michael Morris}.} \bibinfo{year}{2010}\natexlab{}.
\newblock \showarticletitle{Negotiating Gender Roles: Gender Differences in
  Assertive Negotiating Are Mediated by Women's Fear of Backlash and Attenuated
  When Negotiating on Behalf of Others}.
\newblock \bibinfo{journal}{\emph{Journal of personality and social
  psychology}}  \bibinfo{volume}{98} (\bibinfo{year}{2010}),
  \bibinfo{pages}{256--67}.
\newblock
\urldef\tempurl%
\url{https://doi.org/10.1037/a0017094}
\showDOI{\tempurl}


\bibitem[Amanatullah and Tinsley(2013)]%
        {Amanatullah2013AskAY}
\bibfield{author}{\bibinfo{person}{Emily~T. Amanatullah} {and}
  \bibinfo{person}{Catherine~H. Tinsley}.} \bibinfo{year}{2013}\natexlab{}.
\newblock \showarticletitle{Ask and Ye Shall Receive? How Gender and Status
  Moderate Negotiation Success}.
\newblock \bibinfo{journal}{\emph{Negotiation and Conflict Management
  Research}}  \bibinfo{volume}{6} (\bibinfo{year}{2013}),
  \bibinfo{pages}{253--272}.
\newblock
\urldef\tempurl%
\url{https://doi.org/10.1111/ncmr.12014}
\showDOI{\tempurl}


\bibitem[Belk(2013)]%
        {belk2013extended}
\bibfield{author}{\bibinfo{person}{Russell~W. Belk}.}
  \bibinfo{year}{2013}\natexlab{}.
\newblock \showarticletitle{{Extended Self in a Digital World}}.
\newblock \bibinfo{journal}{\emph{Journal of Consumer Research}}
  \bibinfo{volume}{40} (\bibinfo{year}{2013}), \bibinfo{pages}{477--500}.
\newblock
\showISSN{0093-5301}
\urldef\tempurl%
\url{https://doi.org/10.1086/671052}
\showDOI{\tempurl}


\bibitem[Bender et~al\mbox{.}(2021)]%
        {Bender-etal-2021-parrots}
\bibfield{author}{\bibinfo{person}{Emily~M. Bender}, \bibinfo{person}{Timnit
  Gebru}, \bibinfo{person}{Angelina McMillan-Major}, {and}
  \bibinfo{person}{Shmargaret Shmitchell}.} \bibinfo{year}{2021}\natexlab{}.
\newblock \showarticletitle{On the Dangers of Stochastic Parrots: Can Language
  Models Be Too Big?}. In \bibinfo{booktitle}{\emph{Proceedings of the 2021 ACM
  Conference on Fairness, Accountability, and Transparency}}
  \emph{(\bibinfo{series}{FAccT '21})}. \bibinfo{publisher}{Association for
  Computing Machinery}, \bibinfo{address}{New York, NY, USA}.
\newblock
\showISBNx{9781450383097}
\urldef\tempurl%
\url{https://doi.org/10.1145/3442188.3445922}
\showDOI{\tempurl}


\bibitem[Bhat et~al\mbox{.}(2023)]%
        {Bhat-cogpros-2023}
\bibfield{author}{\bibinfo{person}{Advait Bhat}, \bibinfo{person}{Saaket
  Agashe}, \bibinfo{person}{Parth Oberoi}, \bibinfo{person}{Niharika Mohile},
  \bibinfo{person}{Ravi Jangir}, {and} \bibinfo{person}{Anirudha Joshi}.}
  \bibinfo{year}{2023}\natexlab{}.
\newblock \showarticletitle{Interacting with Next-Phrase Suggestions: How
  Suggestion Systems Aid and Influence the Cognitive Processes of Writing}. In
  \bibinfo{booktitle}{\emph{Proceedings of the 28th International Conference on
  Intelligent User Interfaces}} \emph{(\bibinfo{series}{IUI '23})}.
  \bibinfo{publisher}{Association for Computing Machinery},
  \bibinfo{address}{New York, NY, USA}.
\newblock
\showISBNx{9798400701061}
\urldef\tempurl%
\url{https://doi.org/10.1145/3581641.3584060}
\showDOI{\tempurl}


\bibitem[Bird et~al\mbox{.}(2020)]%
        {bird2020fairlearn}
\bibfield{author}{\bibinfo{person}{Sarah Bird}, \bibinfo{person}{Miro Dudík},
  \bibinfo{person}{Richard Edgar}, \bibinfo{person}{Brandon Horn},
  \bibinfo{person}{Roman Lutz}, \bibinfo{person}{Vanessa Milan},
  \bibinfo{person}{Mehrnoosh Sameki}, \bibinfo{person}{Hanna Wallach}, {and}
  \bibinfo{person}{Kathleen Walker}.} \bibinfo{year}{2020}\natexlab{}.
\newblock \bibinfo{booktitle}{\emph{Fairlearn: A toolkit for assessing and
  improving fairness in AI}}.
\newblock \bibinfo{type}{{T}echnical {R}eport} MSR-TR-2020-32.
  \bibinfo{institution}{Microsoft}.
\newblock
\urldef\tempurl%
\url{https://www.microsoft.com/en-us/research/publication/fairlearn-a-toolkit-for-assessing-and-improving-fairness-in-ai/}
\showURL{%
\tempurl}


\bibitem[Braun et~al\mbox{.}(2018)]%
        {braun2018senses}
\bibfield{author}{\bibinfo{person}{Niclas Braun}, \bibinfo{person}{Stefan
  Debener}, \bibinfo{person}{Nadine Spychala}, \bibinfo{person}{Edith
  Bongartz}, \bibinfo{person}{Peter Sörös}, \bibinfo{person}{Helge H.~O.
  Müller}, {and} \bibinfo{person}{Alexandra Philipsen}.}
  \bibinfo{year}{2018}\natexlab{}.
\newblock \showarticletitle{The Senses of Agency and Ownership: A Review}.
\newblock \bibinfo{journal}{\emph{Frontiers in Psychology}}
  \bibinfo{volume}{9} (\bibinfo{year}{2018}).
\newblock
\showISSN{1664-1078}
\urldef\tempurl%
\url{https://doi.org/10.3389/fpsyg.2018.00535}
\showDOI{\tempurl}


\bibitem[Cesafsky et~al\mbox{.}(2019)]%
        {cesafsky2019calibrating}
\bibfield{author}{\bibinfo{person}{Laura Cesafsky}, \bibinfo{person}{Erik
  Stayton}, {and} \bibinfo{person}{Melissa Cefkin}.}
  \bibinfo{year}{2019}\natexlab{}.
\newblock \showarticletitle{Calibrating agency: Human-Autonomy Teaming and the
  future of work amid highly automated systems}. In
  \bibinfo{booktitle}{\emph{Ethnographic Praxis in Industry Conference
  Proceedings}}. Wiley Online Library.
\newblock
\urldef\tempurl%
\url{https://doi.org/10.1111/1559-8918.2019.01265}
\showDOI{\tempurl}


\bibitem[Herring and Martinson(2004)]%
        {Herring-etal-2004-authenticity}
\bibfield{author}{\bibinfo{person}{Susan Herring} {and} \bibinfo{person}{Anna
  Martinson}.} \bibinfo{year}{2004}\natexlab{}.
\newblock \showarticletitle{Assessing Gender Authenticity in Computer-Mediated
  Language UseEvidence From an Identity Game}.
\newblock \bibinfo{journal}{\emph{Journal of Language and Social Psychology - J
  LANG SOC PSYCHOL}}  \bibinfo{volume}{23} (\bibinfo{year}{2004}),
  \bibinfo{pages}{424--446}.
\newblock
\urldef\tempurl%
\url{https://doi.org/10.1177/0261927X04269586}
\showDOI{\tempurl}


\bibitem[Jakesch et~al\mbox{.}(2023)]%
        {jakesch-etal-2023-cowriting}
\bibfield{author}{\bibinfo{person}{Maurice Jakesch}, \bibinfo{person}{Advait
  Bhat}, \bibinfo{person}{Daniel Buschek}, \bibinfo{person}{Lior Zalmanson},
  {and} \bibinfo{person}{Mor Naaman}.} \bibinfo{year}{2023}\natexlab{}.
\newblock \showarticletitle{Co-Writing with Opinionated Language Models Affects
  Users’ Views}. In \bibinfo{booktitle}{\emph{Proceedings of the 2023 CHI
  Conference on Human Factors in Computing Systems}}
  \emph{(\bibinfo{series}{CHI '23})}. \bibinfo{publisher}{Association for
  Computing Machinery}, \bibinfo{address}{New York, NY, USA}.
\newblock
\showISBNx{9781450394215}
\urldef\tempurl%
\url{https://doi.org/10.1145/3544548.3581196}
\showDOI{\tempurl}


\bibitem[Kidd and Birhane(2023)]%
        {kidd-birhane2023}
\bibfield{author}{\bibinfo{person}{Celeste Kidd} {and} \bibinfo{person}{Abeba
  Birhane}.} \bibinfo{year}{2023}\natexlab{}.
\newblock \showarticletitle{How AI can distort human beliefs}.
\newblock \bibinfo{journal}{\emph{Science}}  \bibinfo{volume}{380}
  (\bibinfo{year}{2023}), \bibinfo{pages}{1222--1223}.
\newblock
\urldef\tempurl%
\url{https://doi.org/10.1126/science.adi0248}
\showDOI{\tempurl}


\bibitem[Koenecke et~al\mbox{.}(2020)]%
        {koenecke2020asr}
\bibfield{author}{\bibinfo{person}{Allison Koenecke}, \bibinfo{person}{Andrew
  Nam}, \bibinfo{person}{Emily Lake}, \bibinfo{person}{Joe Nudell},
  \bibinfo{person}{Minnie Quartey}, \bibinfo{person}{Zion Mengesha},
  \bibinfo{person}{Connor Toups}, \bibinfo{person}{John~R. Rickford},
  \bibinfo{person}{Dan Jurafsky}, {and} \bibinfo{person}{Sharad Goel}.}
  \bibinfo{year}{2020}\natexlab{}.
\newblock \showarticletitle{Racial disparities in automated speech
  recognition}.
\newblock \bibinfo{journal}{\emph{Proceedings of the National Academy of
  Sciences}}  \bibinfo{volume}{117} (\bibinfo{year}{2020}),
  \bibinfo{pages}{7684--7689}.
\newblock
\urldef\tempurl%
\url{https://doi.org/10.1073/pnas.1915768117}
\showDOI{\tempurl}


\bibitem[Lee et~al\mbox{.}(2022)]%
        {lee-2022-coauthor}
\bibfield{author}{\bibinfo{person}{Mina Lee}, \bibinfo{person}{Percy Liang},
  {and} \bibinfo{person}{Qian Yang}.} \bibinfo{year}{2022}\natexlab{}.
\newblock \showarticletitle{CoAuthor: Designing a Human-AI Collaborative
  Writing Dataset for Exploring Language Model Capabilities}. In
  \bibinfo{booktitle}{\emph{Proceedings of the 2022 CHI Conference on Human
  Factors in Computing Systems}} \emph{(\bibinfo{series}{CHI '22})}.
  \bibinfo{publisher}{Association for Computing Machinery},
  \bibinfo{address}{New York, NY, USA}.
\newblock
\showISBNx{9781450391573}
\urldef\tempurl%
\url{https://doi.org/10.1145/3491102.3502030}
\showDOI{\tempurl}


\bibitem[Mazei et~al\mbox{.}(2015)]%
        {mazei2015meta}
\bibfield{author}{\bibinfo{person}{Jens Mazei}, \bibinfo{person}{Joachim
  H{\"u}ffmeier}, \bibinfo{person}{Philipp~Alexander Freund},
  \bibinfo{person}{Alice~F Stuhlmacher}, \bibinfo{person}{Lena Bilke}, {and}
  \bibinfo{person}{Guido Hertel}.} \bibinfo{year}{2015}\natexlab{}.
\newblock \showarticletitle{A meta-analysis on gender differences in
  negotiation outcomes and their moderators.}
\newblock \bibinfo{journal}{\emph{Psychological bulletin}}
  \bibinfo{volume}{141} (\bibinfo{year}{2015}), \bibinfo{pages}{85--104}.
\newblock
\urldef\tempurl%
\url{https://doi.org/10.1037/a0038184}
\showDOI{\tempurl}


\bibitem[Mengesha et~al\mbox{.}(2021)]%
        {mengesha2021don}
\bibfield{author}{\bibinfo{person}{Zion Mengesha}, \bibinfo{person}{Courtney
  Heldreth}, \bibinfo{person}{Michal Lahav}, \bibinfo{person}{Juliana
  Sublewski}, {and} \bibinfo{person}{Elyse Tuennerman}.}
  \bibinfo{year}{2021}\natexlab{}.
\newblock \showarticletitle{“I don’t Think These Devices are Very
  Culturally Sensitive.”—Impact of Automated Speech Recognition Errors on
  African Americans}.
\newblock \bibinfo{journal}{\emph{Frontiers in Artificial Intelligence}}
  \bibinfo{volume}{4} (\bibinfo{year}{2021}), \bibinfo{pages}{169}.
\newblock
\urldef\tempurl%
\url{https://doi.org/10.3389/frai.2021.725911}
\showDOI{\tempurl}


\bibitem[Merchant(2012)]%
        {merchant-2012-gender-comm}
\bibfield{author}{\bibinfo{person}{Karima Merchant}.}
  \bibinfo{year}{2012}\natexlab{}.
\newblock \bibinfo{booktitle}{\emph{How Men And Women Differ: Gender
  Differences in Communication Styles, Influence Tactics, and Leadership
  Styles}}.
\newblock \bibinfo{publisher}{Claremont McKenna College}.
\newblock


\bibitem[Mieczkowski et~al\mbox{.}(2021)]%
        {mieczkowski2021ai}
\bibfield{author}{\bibinfo{person}{Hannah Mieczkowski},
  \bibinfo{person}{Jeffrey~T. Hancock}, \bibinfo{person}{Mor Naaman},
  \bibinfo{person}{Malte Jung}, {and} \bibinfo{person}{Jess Hohenstein}.}
  \bibinfo{year}{2021}\natexlab{}.
\newblock \showarticletitle{AI-Mediated Communication: Language Use and
  Interpersonal Effects in a Referential Communication Task}.
\newblock \bibinfo{journal}{\emph{Proc. ACM Hum.-Comput. Interact.}}
  \bibinfo{volume}{5}, \bibinfo{number}{CSCW1}, Article \bibinfo{articleno}{17}
  (\bibinfo{date}{apr} \bibinfo{year}{2021}), \bibinfo{numpages}{14}~pages.
\newblock
\urldef\tempurl%
\url{https://doi.org/10.1145/3449091}
\showDOI{\tempurl}


\bibitem[Mieczkowski(2022)]%
        {mieczkowski2022ai}
\bibfield{author}{\bibinfo{person}{Hannah~N. Mieczkowski}.}
  \bibinfo{year}{2022}\natexlab{}.
\newblock \emph{\bibinfo{title}{AI-Mediated Communication: Examining Agency,
  Ownership, Expertise, and Roles of AI Systems}}.
\newblock \bibinfo{thesistype}{Ph.\,D. Dissertation}.
\newblock
\showISBNx{9798357505163}
\urldef\tempurl%
\url{https://www.proquest.com/dissertations-theses/ai-mediated-communication-examining-agency/docview/2734696274/se-2}
\showURL{%
\tempurl}


\bibitem[Mirowski et~al\mbox{.}(2023)]%
        {mirowski-screenplays-2023}
\bibfield{author}{\bibinfo{person}{Piotr Mirowski}, \bibinfo{person}{Kory~W.
  Mathewson}, \bibinfo{person}{Jaylen Pittman}, {and} \bibinfo{person}{Richard
  Evans}.} \bibinfo{year}{2023}\natexlab{}.
\newblock \showarticletitle{Co-Writing Screenplays and Theatre Scripts with
  Language Models: Evaluation by Industry Professionals}
  \emph{(\bibinfo{series}{CHI '23})}. \bibinfo{publisher}{Association for
  Computing Machinery}, \bibinfo{address}{New York, NY, USA}.
\newblock
\showISBNx{9781450394215}
\urldef\tempurl%
\url{https://doi.org/10.1145/3544548.3581225}
\showDOI{\tempurl}


\bibitem[Mullany(2007)]%
        {mullany-2007-gendered-discourse}
\bibfield{author}{\bibinfo{person}{Louise Mullany}.}
  \bibinfo{year}{2007}\natexlab{}.
\newblock \bibinfo{booktitle}{\emph{Gendered discourse in the professional
  workplace}}.
\newblock \bibinfo{publisher}{Springer}.
\newblock
\urldef\tempurl%
\url{https://doi.org/10.1057/9780230592902}
\showDOI{\tempurl}


\bibitem[Neff and Nagy(2018)]%
        {neff2018agency}
\bibfield{author}{\bibinfo{person}{Gina Neff} {and} \bibinfo{person}{Peter
  Nagy}.} \bibinfo{year}{2018}\natexlab{}.
\newblock \showarticletitle{Agency in the digital age: Using symbiotic agency
  to explain human-technology interaction}.
\newblock In \bibinfo{booktitle}{\emph{A Networked Self and Human Augmentics,
  Artificial Intelligence, Sentience}}, \bibfield{editor}{\bibinfo{person}{Zizi
  Papacharissi}} (Ed.). \bibinfo{publisher}{Routledge}, \bibinfo{pages}{11}.
\newblock
\urldef\tempurl%
\url{https://doi.org/10.4324/9781315202082}
\showDOI{\tempurl}


\bibitem[Noy and Zhang(2023)]%
        {Zhang-2023-productivity}
\bibfield{author}{\bibinfo{person}{Shakked Noy} {and} \bibinfo{person}{Whitney
  Zhang}.} \bibinfo{year}{2023}\natexlab{}.
\newblock \showarticletitle{Experimental evidence on the productivity effects
  of generative artificial intelligence}.
\newblock \bibinfo{journal}{\emph{Science}}  \bibinfo{volume}{381}
  (\bibinfo{year}{2023}), \bibinfo{pages}{187--192}.
\newblock
\urldef\tempurl%
\url{https://doi.org/10.1126/science.adh2586}
\showDOI{\tempurl}


\bibitem[Poddar et~al\mbox{.}(2023)]%
        {poddar2023ai}
\bibfield{author}{\bibinfo{person}{Ritika Poddar}, \bibinfo{person}{Rashmi
  Sinha}, \bibinfo{person}{Mor Naaman}, {and} \bibinfo{person}{Maurice
  Jakesch}.} \bibinfo{year}{2023}\natexlab{}.
\newblock \showarticletitle{AI Writing Assistants Influence Topic Choice in
  Self-Presentation}. In \bibinfo{booktitle}{\emph{Extended Abstracts of the
  2023 CHI Conference on Human Factors in Computing Systems}}
  \emph{(\bibinfo{series}{CHI EA '23})}. \bibinfo{publisher}{Association for
  Computing Machinery}, \bibinfo{address}{New York, NY, USA}.
\newblock
\showISBNx{9781450394222}
\urldef\tempurl%
\url{https://doi.org/10.1145/3544549.3585893}
\showDOI{\tempurl}


\bibitem[Ross and Bayer(2021)]%
        {ross2021explicating}
\bibfield{author}{\bibinfo{person}{Morgan~Quinn Ross} {and}
  \bibinfo{person}{Joseph~B Bayer}.} \bibinfo{year}{2021}\natexlab{}.
\newblock \showarticletitle{Explicating self-phones: Dimensions and correlates
  of smartphone self-extension}.
\newblock \bibinfo{journal}{\emph{Mobile Media \& Communication}}
  \bibinfo{volume}{9} (\bibinfo{year}{2021}), \bibinfo{pages}{488--512}.
\newblock
\urldef\tempurl%
\url{https://doi.org/10.1177/2050157920980508}
\showDOI{\tempurl}


\bibitem[Ryan and Deci(2017)]%
        {ryan2017self}
\bibfield{author}{\bibinfo{person}{Richard~M Ryan} {and}
  \bibinfo{person}{Edward~L Deci}.} \bibinfo{year}{2017}\natexlab{}.
\newblock \bibinfo{booktitle}{\emph{Self-determination theory: Basic
  psychological needs in motivation, development, and wellness}}.
\newblock \bibinfo{publisher}{Guilford publications}.
\newblock


\bibitem[Shelby et~al\mbox{.}(2023)]%
        {shelby2023}
\bibfield{author}{\bibinfo{person}{Renee Shelby}, \bibinfo{person}{Shalaleh
  Rismani}, \bibinfo{person}{Kathryn Henne}, \bibinfo{person}{AJung Moon},
  \bibinfo{person}{Negar Rostamzadeh}, \bibinfo{person}{Paul Nicholas},
  \bibinfo{person}{N'Mah Yilla-Akbari}, \bibinfo{person}{Jess Gallegos},
  \bibinfo{person}{Andrew Smart}, \bibinfo{person}{Emilio Garcia}, {and}
  \bibinfo{person}{Gurleen Virk}.} \bibinfo{year}{2023}\natexlab{}.
\newblock \showarticletitle{Sociotechnical Harms of Algorithmic Systems:
  Scoping a Taxonomy for Harm Reduction}. In
  \bibinfo{booktitle}{\emph{Proceedings of the 2023 AAAI/ACM Conference on AI,
  Ethics, and Society}} \emph{(\bibinfo{series}{AIES '23})}.
  \bibinfo{publisher}{Association for Computing Machinery},
  \bibinfo{address}{New York, NY, USA}.
\newblock
\showISBNx{9798400702310}
\urldef\tempurl%
\url{https://doi.org/10.1145/3600211.3604673}
\showDOI{\tempurl}


\bibitem[Singh et~al\mbox{.}(2023)]%
        {singh-etal-2022-creative}
\bibfield{author}{\bibinfo{person}{Nikhil Singh}, \bibinfo{person}{Guillermo
  Bernal}, \bibinfo{person}{Daria Savchenko}, {and} \bibinfo{person}{Elena~L.
  Glassman}.} \bibinfo{year}{2023}\natexlab{}.
\newblock \showarticletitle{Where to Hide a Stolen Elephant: Leaps in Creative
  Writing with Multimodal Machine Intelligence}.
\newblock \bibinfo{journal}{\emph{ACM Trans. Comput.-Hum. Interact.}}
  \bibinfo{volume}{30}, Article \bibinfo{articleno}{68} (\bibinfo{date}{sep}
  \bibinfo{year}{2023}), \bibinfo{numpages}{57}~pages.
\newblock
\showISSN{1073-0516}
\urldef\tempurl%
\url{https://doi.org/10.1145/3511599}
\showDOI{\tempurl}


\bibitem[Stuhlmacher and Walters(1999)]%
        {stuhlmacher1999gender}
\bibfield{author}{\bibinfo{person}{Alice~F. Stuhlmacher} {and}
  \bibinfo{person}{Amy~E. Walters}.} \bibinfo{year}{1999}\natexlab{}.
\newblock \showarticletitle{Gender differences in negotiation outcome: A
  meta-analysis}.
\newblock \bibinfo{journal}{\emph{Personnel Psychology}}  \bibinfo{volume}{52}
  (\bibinfo{year}{1999}), \bibinfo{pages}{653--677}.
\newblock
\showISBNx{00315826}
\urldef\tempurl%
\url{https://www.proquest.com/scholarly-journals/gender-differences-negotiation-outcome-meta/docview/220142492/se-2}
\showURL{%
\tempurl}


\bibitem[Thimm et~al\mbox{.}(2003)]%
        {thimm-etal-2008-communicating}
\bibfield{author}{\bibinfo{person}{Caja Thimm}, \bibinfo{person}{Sabine~C
  Koch}, {and} \bibinfo{person}{Sabine Schey}.}
  \bibinfo{year}{2003}\natexlab{}.
\newblock \showarticletitle{Communicating gendered professional identity:
  Competence, cooperation, and conflict in the workplace}.
\newblock \bibinfo{journal}{\emph{The handbook of language and gender}}
  (\bibinfo{year}{2003}), \bibinfo{pages}{528--549}.
\newblock
\urldef\tempurl%
\url{https://doi.org/10.1002/9780470756942.ch23}
\showDOI{\tempurl}


\bibitem[Wan et~al\mbox{.}(2023)]%
        {wan2023kelly}
\bibfield{author}{\bibinfo{person}{Yixin Wan}, \bibinfo{person}{George Pu},
  \bibinfo{person}{Jiao Sun}, \bibinfo{person}{Aparna Garimella},
  \bibinfo{person}{Kai-Wei Chang}, {and} \bibinfo{person}{Nanyun Peng}.}
  \bibinfo{year}{2023}\natexlab{}.
\newblock \showarticletitle{{``}Kelly is a Warm Person, Joseph is a Role
  Model{''}: Gender Biases in {LLM}-Generated Reference Letters}. In
  \bibinfo{booktitle}{\emph{Findings of the Association for Computational
  Linguistics: EMNLP 2023}}. \bibinfo{publisher}{Association for Computational
  Linguistics}.
\newblock
\urldef\tempurl%
\url{https://doi.org/10.18653/v1/2023.findings-emnlp.243}
\showDOI{\tempurl}


\bibitem[Weidinger et~al\mbox{.}(2022)]%
        {weidinger-etal-2022-risks}
\bibfield{author}{\bibinfo{person}{Laura Weidinger}, \bibinfo{person}{Jonathan
  Uesato}, \bibinfo{person}{Maribeth Rauh}, \bibinfo{person}{Conor Griffin},
  \bibinfo{person}{Po-Sen Huang}, \bibinfo{person}{John Mellor},
  \bibinfo{person}{Amelia Glaese}, \bibinfo{person}{Myra Cheng},
  \bibinfo{person}{Borja Balle}, \bibinfo{person}{Atoosa Kasirzadeh},
  \bibinfo{person}{Courtney Biles}, \bibinfo{person}{Sasha Brown},
  \bibinfo{person}{Zac Kenton}, \bibinfo{person}{Will Hawkins},
  \bibinfo{person}{Tom Stepleton}, \bibinfo{person}{Abeba Birhane},
  \bibinfo{person}{Lisa~Anne Hendricks}, \bibinfo{person}{Laura Rimell},
  \bibinfo{person}{William Isaac}, \bibinfo{person}{Julia Haas},
  \bibinfo{person}{Sean Legassick}, \bibinfo{person}{Geoffrey Irving}, {and}
  \bibinfo{person}{Iason Gabriel}.} \bibinfo{year}{2022}\natexlab{}.
\newblock \showarticletitle{Taxonomy of Risks posed by Language Models}. In
  \bibinfo{booktitle}{\emph{Proceedings of the 2022 ACM Conference on Fairness,
  Accountability, and Transparency}} \emph{(\bibinfo{series}{FAccT '22})}.
  \bibinfo{publisher}{Association for Computing Machinery},
  \bibinfo{address}{New York, NY, USA}.
\newblock
\showISBNx{9781450393522}
\urldef\tempurl%
\url{https://doi.org/10.1145/3531146.3533088}
\showDOI{\tempurl}


\end{thebibliography}

\appendix
\section{Appendix}

\subsection{Validating the Treatment}
\newtext{We used human annotators and computational methods to validate the effect of the auto-complete suggestions. 
We are able to analyze the auto-complete suggestions as well as the final message, as our writing app logs all auto-complete suggestions seen by the participants.}
\newtext{For our analysis, we randomly sampled 100 participants from each AI treatment condition. We selected suggestions that appeared at the beginning, middle, and end of the writing task for a total of five suggestions per participant. We used LIWC-22 on the auto-complete suggestions to analyze the following features: positive tone, negative tone, pro-social behavior, politeness, conflict, clout, authenticity, and social behavior as a whole. 
Clout refers to the social status or confidence people display in writing.
Authenticity refers to how much a person self-regulates, with low authenticity indicating the person is socially cautious and high authenticity indicating a person is spontaneous.
Based on prior literature~\cite{mullany-2007-gendered-discourse,merchant-2012-gender-comm,thimm-etal-2008-communicating}, self-assured language and hesitant language should differ along these metrics.
With the exception of the positive tone metric, t-tests revealed significant differences between the means of the self-assured model suggestions and hesitant model suggestions across all the metrics (negative tone: t=-6.73, \textit{p}<0.001; pro-social behavior: t=3.29, \textit{p}=0.0012; politeness: t= 3.77, \textit{p}=0.0024, conflict: t=-2.27, \textit{p}=0.025; clout: t=3.15, \textit{p}=0.002; authenticity: t=-4.00, \textit{p}<0.001; social behavior: t=3.24, \textit{p}=0.001). 
Table~\ref{table: suggestions} shows several of the model suggestions with some selected self-assured suggestions highlighted in dark blue and selected hesitant suggestions highlighted in light blue.}

\begin{table*}[b]
\caption{We prompted GPT-4 to generate suggestions in self-assured style and hesitant style. We sampled a few suggestions participants saw while they were writing. Noticeable differences in the suggestions are highlighted. }
\begin{tabular}{@{}p{0.3\linewidth}p{0.7\linewidth}@{}}
\toprule
\textbf{Prompt} & \textbf{Sampled Suggestions} \\ \midrule
Come up with a continuation to the input text in a \textbf{self-assured} manner asking for a raise & 
\makecell[l]{
  ``Be justifiable to review my current compensation package. I believe an \\increase commensurate with my increased...'' \\
  ``Over the past year, I have consistently met and exceeded my performance \\objectives, taking on additional responsibilities without...'' \\
  ``\textcolor{blue}{My compensation should reflect this}. I would appreciate the opportunity \\ to discuss this with you.'' \\
  ``\textcolor{blue}{It is fair to align my compensation with these} \textcolor{blue}{mounting responsibilities}. I have \\ done extensive research on industry.''
} \\ \midrule
Come up with a continuation to the input text in a \textbf{hesitant} manner asking for a raise & 
\makecell[l]{
  ``I certainly don't want to come across as ungrateful or discontented.''\\
  ``...my performance and contributions, if needed. I am not asking for an immediate \\ response, but rather a thoughtful...'' \\
  ``In comparison to the industry standard, my \textcolor{cyan}{current compensation seems a bit...}\\ \textcolor{cyan}{well}, it's not entirely...'' \\
  ``I thought it might be time to discuss the possibility of... \textcolor{cyan}{well, perhaps a slight}\\ \textcolor{cyan}{adjustment} in my...''
} \\ \bottomrule
\label{table: suggestions}
\end{tabular}
\end{table*}

\newtext{Additional analysis showed the suggestions had an effect on the final message. We had three human annotators evaluate all of the final messages on a scale from 1-10, where 1 indicated no assertiveness and 10 indicated high assertiveness. T-tests revealed the means between conditions to be significant (self-assured/control: t=14.56, \textit{p}<0.001; self-assured/hesitant: t= 14.56, \textit{p}<0.001; control/hesitant: t= 10.09, \textit{p}<0.001). 
Using the same LIWC-22 analysis, t-tests showed significant differences between the text produced by these groups, with the exception of the positive tone metric and politeness (negative tone: t=-7.09, \textit{p}<0.001; pro-social behavior: t=4.83, \textit{p}<0.001; conflict: t=-2.24, \textit{p}=0.026; clout: t= 3.00, \textit{p}=0.003; authenticity: t= -2.24, \textit{p}=0.03; social behavior: t= 5.36, \textit{p}<0.001).}


\subsection{Exploratory Analysis Details}
\newtext{
In the exploratory analysis (section \ref{sec:exploratory}), we performed a row-wise average across the inclusion questions and the control and ownership questions to create our agency measure. Table ~\ref{table:correlation_questions} shows the correlation between the post-task questions. The two ownership questions (\emph{message sounds like me}, \emph{message is mine}), two control questions (\emph{control during process}, \emph{control over message}), and two inclusion questions (\emph{assistant made for me}, \emph{assistant sounds like me}) are each highly correlated, as indicated by the bold text.}

\begin{table*}[t]
\centering
\caption{Correlation table for the post-task questions. The highest correlations are between \emph{message is mine} and \emph{message sounds like me} (the two ownership questions), \emph{control during process} and \emph{control over message} (the two control questions), and \emph{assistant made for me} and \emph{assistant sounds like me (the two inclusion questions)}.}
\label{table:correlation_questions}
\resizebox{\textwidth}{!}{%
\begin{tabular}{rrrrrrrr}
  \toprule
  & \thead{message is mine} & \thead{control over message} & \thead{control during process} & \thead{message sounds like me} & \thead{assistant made for me} & \thead{(assistant sounds like me)} & \thead{ai\_reliance} \\ 
  \midrule
  \thead{message is mine} & 1.00 &  &  &  &  &  &  \\ 
  \thead{control over message} & 0.53 & 1.00 &  &  &  &  &  \\ 
  \thead{control during process} & 0.45 & \textbf{0.65} & 1.00 &  &  &  &  \\ 
  \thead{message sounds like me} & \textbf{0.67} & 0.51 & 0.45 & 1.00 &  &  &  \\ 
  \thead{assistant made for me} & 0.23 & 0.21 & 0.21 & 0.37 & 1.00 &  &  \\ 
  \thead{assistant sounds like me} & 0.13 & 0.17 & 0.22 & 0.21 & \textbf{0.56} & 1.00 &  \\ 
  \thead{ai\_reliance} & -0.37 & -0.27 & -0.22 & -0.28 & 0.13 & 0.16 & 1.00 \\ 
  \bottomrule
\end{tabular}%
}
\end{table*}
\newtext{Table~\ref{table:correlation_predictors} shows the correlation analysis for the predictors in our model (Table~\ref{table:regression} in section~\ref{sec:exploratory}). The demographic features \textit{race} and \textit{gender} in the model are binary with 1 corresponding to the privileged group (white, male) and 0 corresponds to minoritized groups (non-white, WNBSD).}

\begin{table*}[h]
\centering
\caption{Correlation table for the predictors in the OLS linear regression} 
\label{table:correlation_predictors}
\begin{tabular}{rrrrrrrr}
  \hline
 & Age & Race & Income & Gender & AI Reliance & Inclusion & AI Style \\ 
  \hline
Age & 1.00 &  &  &  &  &  &  \\ 
  Race & 0.16 & 1.00 &  &  &  &  &  \\ 
  Income & 0.22 & 0.06 & 1.00 &  &  &  &  \\ 
  Gender & -0.02 & -0.05 & 0.08 & 1.00 &  &  &  \\ 
  AI Reliance & 0.04 & -0.01 & 0.09 & 0.07 & 1.00 &  &  \\ 
  Inclusion & -0.08 & -0.12 & -0.08 & 0.02 & 0.17 & 1.00 &  \\ 
  AI Style & -0.00 & 0.03 & 0.04 & -0.10 & 0.14 & 0.10 & 1.00 \\ 
   \hline
\end{tabular}
\end{table*}

\end{document}